\documentclass[12pt]{article}

\usepackage{amsfonts}
\usepackage{amsmath}
\usepackage{amsthm}
\usepackage{cite}
\usepackage{epsfig}
\usepackage{latexsym}
\usepackage{paralist}
\usepackage{fancyhdr}
\usepackage{graphicx}
\numberwithin{equation}{section}
\usepackage[vcentermath]{youngtab}
\usepackage{young}
\usepackage{ytableau}
\usepackage{etex}
\usepackage{braket}
\usepackage{float}
\usepackage{extarrows}
\usepackage{autobreak}

\usepackage{pict2e}   
\usepackage{animate}  
\usepackage{multirow}
\usepackage{bigdelim}
\usepackage{fancybox}
\usepackage{cals}

\def\be{\begin{equation}}
\def\ee{\end{equation}}
\def\bea{\begin{eqnarray}}
\def\eea{\end{eqnarray}}

\renewcommand{\thefootnote}{\fnsymbol{footnote}}

\begin{document}

\hfuzz=100pt
\title{{\Large \bf{Confinement in 3d $\mathcal{N}=2$ exceptional gauge theories} }}
\date{}
\author{ Keita Nii$^a$\footnote{nii@itp.unibe.ch}
%, 
%Yuta Sekiguchi$^a$\footnote{yuta@itp.unibe.ch}
% and others$^{c,d}$
}
\date{\today}

\maketitle

\thispagestyle{fancy}
%\rhead{****-**-**}
\cfoot{}
\renewcommand{\headrulewidth}{0.0pt}

\vspace*{-1cm}
\begin{center}
%  \spa{0.5} \\
$^{a}${{\it Albert Einstein Center for Fundamental Physics }}
\\{{\it Institute for Theoretical Physics
}}
\\ {{\it University of Bern}}  
\\{{\it  Sidlerstrasse 5, CH-3012 Bern, Switzerland}}
%\\ {{\it  }}
 % \spa{0.5} \\
%$^b${{\it Department of Physics}}
%\\ {{\it Nagoya University, Nagoya 464-8602, Japan}}
%\spa{0.5}  \\
%$^c${{\it  }}
%\\ {{\it }}
%\spa{0.5}  \\
%$^d${{\it }}

\end{center}

\begin{abstract}
We study the low-energy dynamics in three-dimensional $\mathcal{N}=2$ exceptional gauge theories with matters in a fundamental representation, especially focusing on confinement phases and on a quantum structure of the Coulomb branch in the moduli space of vacua. We argue that the confinement phases of these exceptional gauge theories have a single Coulomb branch. The 3d s-confinement phases for the exceptional gauge groups are associated with quantum-deformed moduli spaces of the corresponding 4d $\mathcal{N}=1$ exceptional gauge theories. 
\end{abstract}

\renewcommand{\thefootnote}{\arabic{footnote}}
\setcounter{footnote}{0}

\newpage
\tableofcontents 
\clearpage

\newpage

%%%%%%%%%%%%%%%%%%%%%%%%%%%%%%%%%%%%%%%%%%%%%%%%%%%%%%%%%%
%%%%%%%%%%%%%%%%%%%%%%%%%%%%%%%%%%%%%%%%%%%%%%%%%%%%%%%%%%
\section{Introduction}
%%%%%%%%%%%%%%%%%%%%%%%%%%%%%%%%%%%%%%%%%%%%%%%%%%%%%%%%%%
%%%%%%%%%%%%%%%%%%%%%%%%%%%%%%%%%%%%%%%%%%%%%%%%%%%%%%%%%%
%[SUSY]
Supersymmetry allows us to exactly capture low-energy behaviors of strongly-coupled and non-perturbative gauge theories by using various non-renormalization theorems and holomorphy \cite{Seiberg:1994bz}. Depending on the number of dynamical quarks, the theory exhibits supersymmetry breaking, confinement, non-abelian Coulomb phases and so on. For a non-abelian Coulomb phase, Seiberg duality \cite{Seiberg:1994pq} claims that the low-energy dynamics is equivalent to a magnetic dual theory with a different gauge group.

%[s-confinement]
In this paper, we are interested in s-confinement phases of the 3d $\mathcal{N}=2$ supersymmetric gauge theories. The s-confinement is a confining phase without symmetry breaking at the origin of the moduli space \cite{Seiberg:1994bz, Csaki:1996sm, Csaki:1996zb}. The dual description becomes a non-gauge theory with gauge-singlet chiral superfields. From a viewpoint of the Seiberg duality, s-confinement corresponds to the case with vanishing magnetic gauge groups. After the Seiberg duality was first proposed in a simplest case \cite{Seiberg:1994bz}, many similar dualities were found in 4d \cite{Intriligator:1995id, Intriligator:1995ne, Pouliot:1995zc, Kutasov:1995ve, Kutasov:1995np, Intriligator:1995ax}. However, these dualities are not all generalized to 3d dualities\footnote{For 3d Seiberg dualities, see for example, \cite{Karch:1997ux, Aharony:1997gp, Giveon:2008zn, Niarchos:2008jb, Benini:2011mf, Kapustin:2011vz, Aharony:2013dha, Aharony:2013kma}.}.
This is because the 3d theory has a Coulomb moduli space from the vector superfield, which is extremely modified from the classical picture due to quantum and non-perturbative effects. The lack of this understanding makes the construction of the 3d Seiberg duality very difficult. In this paper, we will give a quantum structure of the Coulomb branch of the 3d $\mathcal{N}=2$ exceptional gauge theories, paying attention to the s-confining phases. The corresponding 4d theories were studied in the literature \cite{Ramond:1996ku, Distler:1996ub, Karch:1997jp, Cho:1997am, Grinstein:1998bu, Pouliot:2001iw}.

%[index constraint]
In four dimensional cases, the quantum low-energy phases are classified by the index constraint of the dynamical quarks \cite{Csaki:1996sm, Csaki:1996zb, Grinstein:1997zv, Grinstein:1998bu}. In order to have an s-confinement phase, the moduli fields cannot have any singularity at the origin of the moduli space. In addition, the gauge singlets (confined degrees of freedom) must be constrained by effective superpotentials in order to reduce the number of the independent Higgs branch operators, which results in a certain index constraints on the matter content
\begin{align}
T_2 (\mathbf{Adj.}) -\sum_i T_2 ( \mathbf{r}_i)=2,
\end{align}
where $T_2 ( \mathbf{r}_i)$ represents the Dynkin index of the representation $\mathbf{r}_i$ for the dynamical matter. The summation is taken over all the chiral superfields coupled with the vector multiplet. 
For phases with a quantum deformed moduli space to be possible, r-charges of certain products of the moduli coordinates should be canceled. This is possible only when the matter content satisfies a similar index constraint.
\begin{align}
T_2 (\mathbf{Adj.}) -\sum_i T_2 ( \mathbf{r}_i)=0, \label{3dscon_ind}
\end{align}

In three dimensional cases, a similar index constraint was obtained in \cite{Intriligator:2013lca, Witten:1999ds} for Chern-Simons-matter theories by requiring that the Witten index (see \cite{Witten:1982df, Smilga:2013usa}) should be unity. This results in the following index constraint
\begin{align}
T_2 (\mathbf{Adj.}) -\sum_i T_2 ( \mathbf{r}_i)=|k|,
\end{align}
where $k$ is a Chern-Simons level. Furthermore, as noticed in \cite{Aharony:1997bx, Csaki:2014cwa, Amariti:2015kha, Nii:2016jzi}, the 3d s-confinement phases are related to the 4d quantum-deformed moduli space via the dimensional reduction procedure. Therefore, it is natural to expect that the 3d s-confinement could appear from the 3d Yang-Mills theories with the index constraint \eqref{3dscon_ind}.
As is well known, these index constraints are not sufficient but necessary conditions. Hence, it would be very important to examine whether or not the theories satisfying the index constraint actually exhibit s-confinement. 
%[In this paper]
In this paper, we investigate the 3d $\mathcal{N}=2$ exceptional gauge theories with fundamental matters, especially focusing on the s-confinement phases with the constraint \eqref{3dscon_ind}. The 3d $G_2$ case was already studied by the author \cite{Nii:2017npz}.

%[organization]
The rest of this paper is organized as follows.
In Section 2, we will review the low-energy dynamics of the 3d $\mathcal{N}=2$ $G_2$ gauge theory with fundamental matters, which illustrates how the classical Coulomb branch in the moduli space of vacua is truncated and modified. We discuss what kind of Coulomb branch can remain exactly massless and stable. 
From Section 3 to Section 5, we will study the 3d $\mathcal{N}=2$ $F_4$, $E_6$ and $E_7$ gauge theories with fundamental matters by focusing on the s-confinement phases. For the $E_6$ case, we can consider ``chiral'' and ``vector-like'' theories by changing the numbers of fundamental and anti-fundamental matters. 
In Section 6, we summarize these results and discuss future directions.

%%%%%%%%%%%%%%%%%%%%%%%%%%%%%%%%%%%%%%%%%%%%%%%%%%%%%%%%%%
%%%%%%%%%%%%%%%%%%%%%%%%%%%%%%%%%%%%%%%%%%%%%%%%%%%%%%%%%%
\section{3d $\mathcal{N}=2$ $G_2$ gauge theory}
%%%%%%%%%%%%%%%%%%%%%%%%%%%%%%%%%%%%%%%%%%%%%%%%%%%%%%%%%%
%%%%%%%%%%%%%%%%%%%%%%%%%%%%%%%%%%%%%%%%%%%%%%%%%%%%%%%%%%
In this section, we briefly review the low-energy dynamics of the 3d $\mathcal{N}=2$ $G_2$ gauge theory with $F$ fundamental matters. This was studied in \cite{Nii:2017npz} whereas the 4d $\mathcal{N}=1$ $G_2$ gauge theory was studied in \cite{Pesando:1995bq, Giddings:1995ns, Pouliot:1995zc, Pouliot:2001iw}. For the analysis of the Higgs branch which is the same as the 4d one, see \cite{Pesando:1995bq, Giddings:1995ns, Pouliot:1995zc, Pouliot:2001iw}. The elementary fields and their quantum numbers are summarized in Table \ref{G2F}. As opposed to the 4d case, the global symmetry of the theory is $SU(F) \times U(1) \times U(1)_R$ since there is no chiral anomaly.

\begin{table}[H]\caption{3d $\mathcal{N}=2$ $G_2$ gauge theory with $F$ fundamental matters} 
\begin{center}
\scalebox{1}{
  \begin{tabular}{|c||c|c|c|c| } \hline
  &$F_4$&$SU(F)$&$U(1)$&$U(1)_R$  \\ \hline
$Q$&$\mathbf{7}$&${\tiny \yng(1)}$&1&$r$ \\ \hline
$M:=QQ$&1&${\tiny \yng(2)}$&2&$2r$ \\
$B:=Q^3$&1&${\tiny \yng(1,1,1)}$&3&$3r$  \\[5pt]
$F:=Q^4$&1&${\tiny \yng(1,1,1,1)}$&4&$4r$ \\[7pt] \hline
$Z_{SU(2)}$&1&1&$-2F$&$2F-6 -2Fr$  \\ \hline
  \end{tabular}}
  \end{center}\label{G2F}
\end{table}

Classically, there are various Coulomb branches which are accessible by tuning the vacuum expectation values of the adjoint scalar in a $G_2$ vector superfield. These are described by combinations of two monopole operators corresponding to two simple roots in the $G_2$ algebra. Along these Coulomb branches, by dualizing the resulting $U(1)$ vector superfield, we can obtain the Coulomb branch coordinates which are the chiral superfields. In the $G_2$ case, we can see two breaking patterns\footnote{For branching rules appearing in this paper, see for example \cite{Slansky:1981yr, Feger:2012bs}.} where the $SU(2)$ subgroup is unbroken
\begin{align}
G_2 & \rightarrow su(2) \times u(1)  \nonumber \\
\mathbf{7} & \rightarrow \mathbf{2}_{\pm 1} +\mathbf{1}_{\pm 2} +\mathbf{1}_0 \\
\mathbf{14} & \rightarrow \mathbf{3}_0+\mathbf{1}_0+ \mathbf{2}_{\pm 3} +\mathbf{2}_{\pm 1} +\mathbf{1}_{\pm2},   \\
G_2 & \rightarrow su(2) \times u(1) \nonumber \\
\mathbf{7} & \rightarrow   \mathbf{2}_{\pm 1}+ \mathbf{3}_{0}  \\
\mathbf{14} &  \rightarrow    \mathbf{3}_{0}+ \mathbf{1}_{0} + \mathbf{1}_{\pm  2}+ \mathbf{4}_{\pm 1}.
\end{align}
Quantum-mechanically, the first Coulomb branch is not allowed since the low-energy $SU(2)$ gauge theory includes no massless dynamical quark and since its vacuum is unstable due to the Affleck-Harvey-Witten superpotential \cite{Affleck:1982as, Aharony:1997bx, deBoer:1997kr}. Therefore, the first Coulomb branch is removed from the quantum moduli space although it is classically allowed. On the other hand, the second Coulomb branch is quantum-mechanically flat and supersymmetric since the massless component $ \mathbf{3}_{0}$ from the fundamental matter allows a supersymmetric and stable vacuum. As a result, we need to introduce a single coordinate $Z_{SU(2)}$ to parametrize the second Coulomb branch. For a more detailed analysis of the $G_2$ Coulomb branch, see \cite{Nii:2017npz}.

From Table \ref{G2F}, we can write down non-singular effective superpotentials for the range $1 \le F \le 4$ as follows
\begin{gather}
W_{F=1} =  \left( \frac{1}{Z_{SU(2)} M} \right)^{\frac{1}{2}},~~~~~
W_{F=2} =   \frac{1}{Z_{SU(2)} \det \, M }  \\
W_{F=3} = \lambda \left[ Z_{SU(2)}(\det \, M-B^2) -1\right],~~~
W_{F=4} =Z_{SU(2)} \left( -\det \,  M +F^2 +BMB  \right),
\end{gather}
where $\lambda$ is a Lagrange multiplier field. 
For $F=4$ where \eqref{3dscon_ind} is satisfied, the theory is in an s-confinement phase and the low-energy dynamics is described by the gauge singlet chiral superfields $M, B, F$ and $Z_{SU(2)}$. Importantly, the origin of the moduli space is not singular and realizes the confinement without symmetry breaking. The F-flatness condition of $Z_{SU(2)}$ leads to $\det \,  M =F^2 +BMB$ which realizes a classical syzygy between the Higgs branch operators \cite{Pouliot:2001iw}. 
For $F=3$, the low-energy theory is described by $M, B$ and $Z_{SU(2)}$ although these are related by one constraint $Z_{SU(2)}(\det \, M-B^2) =1$. Therefore, the origin of the moduli space is removed. Therefore, the confining phase with $F=3$ is necessarily accompanied by global symmetry breaking. For $F \le 2$, there is no stable SUSY vacuum. In the following sections, we will be interested in the s-confinement phases for other exceptional gauge groups.

%%%%%%%%%%%%%%%%%%%%%%%%%%%%%%%%%%%%%%%%%%%%%%%%%%%%%%%%%%
%%%%%%%%%%%%%%%%%%%%%%%%%%%%%%%%%%%%%%%%%%%%%%%%%%%%%%%%%%
\section{3d $\mathcal{N}=2$ $F_4$ gauge theory}
%%%%%%%%%%%%%%%%%%%%%%%%%%%%%%%%%%%%%%%%%%%%%%%%%%%%%%%%%%
%%%%%%%%%%%%%%%%%%%%%%%%%%%%%%%%%%%%%%%%%%%%%%%%%%%%%%%%%%
We start with the s-confinement phase in the 3d $\mathcal{N}=2$ $F_4$ gauge theory with $F$ fundamental matters. The dimension of the fundamental representation is $\mathbf{26}$ whereas the adjoint representation is denoted by $\mathbf{52}$. The Coulomb branch of the 3d $\mathcal{N}=2$ pure $F_4$ SYM theory without a matter was studied in \cite{Davies:2000nw} which focuses on the relation between the 3d and 4d theories. The low-energy dynamics of the 4d $\mathcal{N}=1$ $F_4$ gauge theory with $F$ fundamental matters was studied in \cite{Grinstein:1998bu, Cho:1997am, Pouliot:2001iw}. The Higgs branch in the 3d theory is the same as the 4d one and it is summarized in Table \ref{F4three}.

\begin{table}[H]\caption{Fermion zero-modes for the $F_4$ fundamental monopoles} 
\begin{center}
  \begin{tabular}{|c||c|c| } \hline
  &adjoint& fundamental  \\ \hline
$Y_1$&2&0  \\
$Y_2$&2&0 \\
$Y_3$&2&2 \\
$Y_4$&2&2  \\ \hline 
$Z_{USp(6)}:=Y_1^2 Y_2^3 Y_3^2 Y_4$&$16$&$6$  \\ \hline
  \end{tabular}
  \end{center}\label{zeromodeF4}
\end{table}

Let us consider the Coulomb branch in the 3d $\mathcal{N}=2$ $F_4$ gauge theory with $F$ fundamental matters. Classically, the dimension of the Coulomb branch is four and described by four (fundamental) monopole operators corresponding to the simple roots in the $F_4$ algebra. However, the most of the classical Coulomb branches are not allowed since the vacuum of the unbroken gauge group is non-perturbatively unstable and non-supersymmetric. Table \ref{zeromodeF4} shows the numbers of the fermion zero-modes for the fundamental monopoles. The zero-modes of the adjoint fermion come from the gaugino field in the vector superfield.  
Since $Y_1$ and $Y_2$ only have two gaugino zero-modes, these monopoles can create a non-perturbative superpotential \cite{Affleck:1982as}
\begin{align}
W_{eff}= \frac{1}{Y_1} +\frac{1}{Y_2}. \label{monopoleWF}
\end{align}
For the pure $F_4$ SYM theory without a matter, the monopole generates a similar potential for $Y_3$ and $Y_4$ as well and lifts all the Coulomb branches. As a result, the 3d $\mathcal{N}=2$ pure $F_4$ SYM theory is not supersymmetric \cite{Davies:2000nw}.
For the theory with fundamental matters, it would be natural to think that the Coulomb branch becomes two-dimensional due to \eqref{monopoleWF}. In the following discussion, we show that there are indeed two Coulomb branch operators and discuss their stability. 

The first Coulomb branch denoted by $Z_{USp(6)}$ corresponds to the gauge symmetry breaking
\begin{align}
F_4 & \rightarrow usp(6) \times u(1) \\
\mathbf{26} & \rightarrow \mathbf{14}_0  + \mathbf{6}_{\pm 1}  \\
 \mathbf{52} & \rightarrow  \mathbf{21}_0 + \mathbf{14'}_{\pm 1} + \mathbf{1}_{\pm 2} + \mathbf{1}_0
\end{align}
where $\mathbf{14}$ is a trace-less anti-symmetric tensor and $\mathbf{21}$ is an adjoint (symmetric) tensor in the unbroken $USp(6)$ subgroup. The Coulomb branch operator $Z_{USp(6)}$ corresponds to the dualized chiral superfield constructed from the low-energy $U(1)$ vector superfield. The components with non-zero $U(1)$ charges are all massive and integrated-out. This leads to the mixed Chern-Simons term between the $U(1)$ gauge and global symmetries
\begin{align}
k_{eff}^{U(1), Q} =6FQ_{\mathbf{26}} +16 Q_{\lambda},
\end{align}
where $Q_{\mathbf{26}}$ is a global charge of the fundamental matters and $Q_{\lambda}$ is a global charge of the gaugino field. This operator can be exactly massless since the low-energy $USp(6)$ gauge theory has massless dynamical matters $\mathbf{14}_0$ and its vacuum is made stable and supersymmetric.

The second candidate of the Coulomb branch operators that can survive quantum corrections corresponds to the gauge symmetry breaking 
\begin{align}
F_4 & \rightarrow so(7) \times u(1) \\ 
\mathbf{26} & \rightarrow  \mathbf{7}_0 + \mathbf{1}_{\pm 2} + \mathbf{8}_{\pm 1} + \mathbf{1}_0 \\
 \mathbf{52} & \rightarrow  \mathbf{21}_0 +\mathbf{1}_0+ \mathbf{7}_{\pm 2} + \mathbf{8}_{\pm 1}.
\end{align}
We denote this operator as $Z_{SO(7)}$. Along this branch, the fundamental matter reduces to the massless vector representation $\mathbf{7}_0$ which can make the vacuum of the low-energy $SO(7)$ theory stable and supersymmetric. This is possible only for $F \ge 5$ \cite{Nii:2018tnd}. In this paper, we are interested in confinement phases and we will find that the s-confinement occurs in $F=3$. Therefore, we don't consider the $SO(7)$ direction of the Coulomb branch in what follows. The zero-modes for these two monopole operators are listed in Table \ref{zeromodeF4CB}.

\begin{table}[H]\caption{Fermion zero-modes of the Coulomb branch operators} 
\begin{center}
  \begin{tabular}{|c||c|c| } \hline
  &adjoint& fundamental  \\ \hline
$Z_{USp(6)}$&$16$&$6$  \\ 
$Z_{SO(7)}$&$22$&$10$  \\ \hline
  \end{tabular}
  \end{center}\label{zeromodeF4CB}
\end{table}

The s-confinement phase appears in the 3d $\mathcal{N}=2$ $F_4$ gauge theory with three fundamental matters, where the index constraint \eqref{3dscon_ind} is satisfied. The Higgs branch is studied in four dimensions \cite{Grinstein:1998bu, Cho:1997am, Pouliot:2001iw} and described by six composite operators. From Table \ref{F4three}, we can write down a effective superpotential as follows.
\begin{align}
W& = Z_{USp(6)} \biggl(N_9^2 +N_5^2N_4^2 +N_5^2 M_2^4 +N_5^2 N_4 M_2^2 +N_5M_2^5 M_3 +N_5 M_2^3N_4M_3 +N_5 M_2N_4^2M_3 \nonumber \\
&\quad +  N_4^3N_6 +N_4^3 M_3^2+ N_4^2 M_2^2 N_6 +N_4^2M_2^2 M_3^2 +N_4 M_2^4 N_6 + N_4M_2^4M_3^2 +M_2^6 N_6 +M_2^6M_3^2\biggr) \label{3dF4W}
\end{align}
The F-flatness condition for $Z_{USp(6)}$ imposes one constraint between the Higgs branch operators and this is consistent with the classical Higgs branch: At a generic point of the Higgs branch, the gauge group is completely higgsed. Therefore, the $2 \times \mathbf{26}$ components of the three fundamental matters are eaten via the Higgs mechanism. The remaining components which explain the classical dimensions of the Higgs branch is $26$-dimensional. The total number of the Higgs branch operators introduced in Table \ref{F4three} is $27$ and the F-flatness condition for $Z_{USp(6)}$ correctly reduces them by one.  

\begin{table}[H]\caption{3d $\mathcal{N}=2$ $F_4$ gauge theory with $F=3$ fundamental matters} 
\begin{center}
\scalebox{1}{
  \begin{tabular}{|c||c|c|c|c| } \hline
  &$F_4$&$SU(F=3)$&$U(1)$&$U(1)_R$  \\ \hline
$Q$&$\mathbf{26}$&${\tiny \yng(1)}$&1&$r$ \\
$\lambda$&$\mathbf{52}$&1&0&$1$ \\
$\eta=\Lambda^b$&1&1&$18$&$6F(r -1) +18=18r$  \\  \hline
$M_2:=Q^2$&1&${\tiny \yng(2)}$&2&$2r$ \\
$M_3:=Q^3$&1&${\tiny \yng(3)}$&3&$3r$ \\
$N_{4}:=Q^4$&1&${\tiny \yng(2,2)} = {\tiny \overline{\yng(2)}}$&4&$4r$ \\[5pt]
$N_{5}:=Q^5$&1&${\tiny \yng(2,2,1)} = {\tiny \overline{\yng(1)}}$&5&$5r$ \\[5pt]
$N_{6}:=Q^6$&1&${\tiny \yng(2,2,2)} =1$&6& $6r$ \\[5pt]
$N_{9}:=Q^9$&1&${\tiny \yng(3,3,3)} =1$&9& $9r$  \\[5pt] \hline  
$Z_{USp(6)}$&1&1&$-6F=-18$& $-6F(r-1) -16=2-18r$  \\ \hline
  \end{tabular}}
  \end{center}\label{F4three}
\end{table}

Let us test the validity of the above s-confinement phase by flowing to the Higgs branch where the low-energy theory becomes a 3d $\mathcal{N}=2$ $Spin(9)$ theory with three vectors and two spinors. This can be achieved by introducing a rank-$1$ vacuum expectation value to $\braket{M_2^{~33}}:=\braket{\delta^{ab}Q_a^{~3}Q_b^{~3}}=v^2$. Along this breaking, the fundamental matter is decomposed as $\mathbf{26} \rightarrow \mathbf{1}+\mathbf{9}+\mathbf{16}$ and a single $\mathbf{16}$ is eaten via the Higgs mechanism. 
 In \cite{Nii:2018wwj}, it was shown that the 3d $\mathcal{N}=2$ $Spin(9)$ theory with three vectors and two spinors exhibits an s-confinement phase with a single Coulomb branch. The Coulomb branch of the low-energy $Spin(9)$ theory has $14$ matter fermion zero-modes from vector and spinor matters. The low-energy Coulomb branch is identified with $Z_{USp(6)}v^4$, where $v^4$ is absorbing four fermion zero-modes. This flow is consistent with our analysis of the $F_4$ s-confinement phase.

Next, we consider a different Higgs branch along which the $SO(8)$ gauge symmetry remains unbroken. This can be achieved by introducing a non-zero expectation value to a fundamental matter such that $M_3:=d^{abc}Q_aQ_bQ_c$ is non-zero. The fundamental representation is decomposed as $\mathbf{26} \rightarrow \mathbf{8_v}+\mathbf{8_s} +\mathbf{8_c}+\mathbf{1}+\mathbf{1}$. The low-energy theory becomes a 3d $\mathcal{N}=2$ $Spin(8)$ gauge theory with two vectors, two spinors and two conjugate spinors. This theory was recently studied in \cite{Nii:2018wwj} and its low-energy theory exhibits s-confinement. The Coulomb branch of the $Spin(8)$ theory is one-dimensional and  the monopole associated with this operator has $12$ fermion zero-modes from the matter fields. This is consistent with the Coulomb branch for the $F_4$ theory since the two fundamental $\mathbf{26}$ matters (one fundamental matter is eaten via the Higgs mechanism.) have $12$ fermion-zero modes.

We can connect the 3d and 4d dynamics by putting the 4d $\mathcal{N}=1$ $F_4$ gauge theory with three fundamental matters on a circle. When the 4d theory is defined in $\mathbb{S}^1 \times \mathbb{R}^3$, there is an additional monopole configuration known as a KK-monopole \cite{Lee:1997vp, Lee:1998vu, Aharony:1997bx, Davies:2000nw}, which is a twisted instanton in $\mathbb{S}^1 \times \mathbb{R}^3$. The KK-monopole generates a non-perturbative superpotential
\begin{align}
W_{\mathbb{S}^1 \times \mathbb{R}^3} = \eta Z_{USp(6)}, \label{F4WKK}
\end{align}
where $\eta$ is a dynamical scale of the 4d gauge interaction. By combining the two effective superpotentials \eqref{3dF4W} and \eqref{F4WKK}, and by integrating out the Coulomb branch which is absent in a 4d limit, we can reproduce the 4d quantum-deformed constraint on the moduli space \cite{Cho:1997am, Grinstein:1998bu}. This serves as another test of our analysis.

The similar analysis based on the Coulomb branch $Z_{USp(6)}$ is possible for small flavors $F \le 2$. For $F=1$ and $F=2$, we find that the symmetry argument is consistent with the following effective superpotentials
\begin{align}
W_{F=1}  &= \left[ \frac{1}{Z_{USp(6)}  (M_2^3+ M_3^2) } \right]^{\frac{1}{5}}  \label{WF4f1}\\
W_{F=2}  &= \left[ \frac{1}{Z_{USp(6)}  (N_4^3 +N_4(M_2^4+M_2M_3^2) +M_2^6+M_2^3M_3^2+M_3^4)  } \right]^{\frac{1}{2}}.   \label{WF4f2}
\end{align}
These are runaway superpotentials and we can conclude that there is no stable supersymmetric vacuum for $F=1,2$.

%%%%%%%%%%%%%%%%%%%%%%%%%%%%%%%%%%%%%%%%%%%%%%%%%%%%%%%%%%
%%%%%%%%%%%%%%%%%%%%%%%%%%%%%%%%%%%%%%%%%%%%%%%%%%%%%%%%%%
\section{3d $\mathcal{N}=2$ $E_6$ gauge theory}
%%%%%%%%%%%%%%%%%%%%%%%%%%%%%%%%%%%%%%%%%%%%%%%%%%%%%%%%%%
%%%%%%%%%%%%%%%%%%%%%%%%%%%%%%%%%%%%%%%%%%%%%%%%%%%%%%%%%%
Let us move on to the 3d $\mathcal{N}=2$ $E_6$ gauge theory with dynamical matters in (anti-)fundamental representations. Again, we mostly restrict our attention to the s-confinement phase. The corresponding 4d theory was studied in \cite{Cho:1997am, Grinstein:1998bu, Pouliot:2001iw, Ramond:1996ku, Distler:1996ub, Karch:1997jp}. The structure of the Higgs branch is the same as the 4d one. In 4d, there are the three examples which exhibit a quantum-deformed moduli space \cite{Cho:1997am, Grinstein:1998bu} with the index constraint \eqref{3dscon_ind} satisfied. Corresponding to these cases, we will find the s-confinement descriptions of the 3d $E_6$ gauge theories.

Since we are studying the 3d theory, there is a Coulomb branch of the adjoint scalar from the vector superfield. When the Coulomb branch operator $Z_{SU(6)}$ obtains a non-zero expectation value, the gauge group is spontaneously broken to
\begin{align}
E_6  & \rightarrow su(6) \times u(1) \\
\mathbf{27} &  \rightarrow \mathbf{15}_0 + \overline{\mathbf{6}}_{\pm 1} \\
\overline{\mathbf{27}} &  \rightarrow \overline{\mathbf{15}}_0 + \mathbf{6}_{\pm 1} \\
\mathbf{78} &  \rightarrow  \mathbf{35}_0 + \mathbf{1}_{0}+\mathbf{1}_{\pm 2} +\mathbf{20}_{\pm 1}.
\end{align}
This operator was studied also in the 4d $\mathcal{N}=1$ $E_6$ pure SYM theory defined on a circle \cite{Davies:2000nw}.
The massless components $\mathbf{15}_0$ and $ \overline{\mathbf{15}}_0$ can make the vacuum of the low-energy $SU(6)$ gauge theory stable and supersymmetric. Therefore, this branch could be quantum-mechanically stable and exactly massless. For a pure $E_6$ SYM theory without a matter, the low-energy $SU(6)$ pure SYM theory generates a runaway potential \cite{Affleck:1982as, Aharony:1997bx} and there is no stable Coulomb brach. The massive components are integrated out and lead to the mixed Chern-Simons terms which give rise to the non-trivial charges of the Coulomb branch operator \cite{Intriligator:2013lca}. The zero-modes of the monopole corresponding to $Z_{SU(6)}$ is summarized in Table \ref{zeromodeE6CB}. Based on this Coulomb brach, we find three examples of s-confinement. 

\begin{table}[H]\caption{Fermion zero-modes of the $E_6$ Coulomb branch operator $Z_{SU(6)}$} 
\begin{center}
  \begin{tabular}{|c||c|c|c| } \hline
  &adjoint& fundamental & anti-fundamental  \\ \hline
$Z_{SU(6)}$&$22$&$6$ &$6$ \\ \hline
  \end{tabular}
  \end{center}\label{zeromodeE6CB}
\end{table}

%%%%%%%%%%%%%%%%%%%%%%%%%%%%%%%%%%%%%%%%%%%%%%%%%%%%%%%%%%
\subsection{$E_6$ with four fundamental matters}
%%%%%%%%%%%%%%%%%%%%%%%%%%%%%%%%%%%%%%%%%%%%%%%%%%%%%%%%%%
The first example is a 3d $\mathcal{N}=2$ $E_6$ gauge theory with four fundamental matters. The quantum numbers of the Higgs and Coulomb branch operators are summarized in Table \ref{E640}. The effective superpotential becomes 
\begin{align}
W_{N_f=4}= Z_{SU(6)} \left[ D^2 +C^4+C^3B^2+C^2B^4+CB^6+B^8  \right], \label{WsupE640}
\end{align}
which is consistent with all the symmetries. 

By giving an expectation value to $\braket{B}:=\braket{f_{abc}Q^aQ^b Q^c}=v^3$, the gauge group is higgsed to $F_4$. The matter fields are decomposed into $\mathbf{27}  \rightarrow \mathbf{26}+\mathbf{1}$ and a single fundamental matter is eaten via the Higgs mechanism. Hence, the low-energy description becomes a 3d $\mathcal{N}=2$ $F_4$ gauge theory with three fundamental matters, which was studied in the previous section and exhibits an s-confinement phase. In the dual (confining) description, $D$ is identified with $v^3 N_9$ while $C$ is decomposed into $N_6, v N_5$ and $v^2 N_4$. The cubic composite $B$ reduces to $vM_2$ and $M_3$. By inserting them into \eqref{WsupE640}, the $F_4$ superpotential \eqref{3dF4W} is re-derived. The high- and low-energy Coulomb branches are identified as $Z_{SU(6)} v^6 \sim Z_{USp(6)} $.

When the 4d $\mathcal{N}=2$ $E_6$ gauge theory is put on a circle, the twisted instanton, which is known also as a KK-monopole, generates a non-perturbative superpotential
\begin{align}
W= \eta Z_{SU(6)}, \label{WsupKKE6}
\end{align}
where $\eta$ is a dynamical scale of the 4d gauge interaction. In the 4d limit (de-compactification limit), the Coulomb brach is massive and should be integrated out. By combining \eqref{WsupE640} and \eqref{WsupKKE6}, we can reproduce the 4d quantum-deformed moduli space \cite{Cho:1997am, Grinstein:1998bu}. This is another consistency check of our analysis. 

 We can also study the low-energy dynamics of the 3d $\mathcal{N}=2$ $E_6$ gauge theory with $F <4$ fundamental matters by using the Coulomb branch $Z_{SU(6)}$. For $F=1$ and $F=2$, the Higgs branch is described only by $B=Q^3$. For $F=3$, the Higgs branch is described by $B=Q^3$ and $C=Q^6$. For these cases, the effective superpotentials become 
\begin{gather}
W_{F=1} = \left( \frac{1}{Z_{SU(6)} B^2} \right)^{\frac{1}{8}},~~~~
W_{F=2} = \left( \frac{1}{Z_{SU(6)} B^4} \right)^{\frac{1}{5}} \\
W_{F=3} = \left[ \frac{1}{Z_{SU(6)} (B^6 +B^4 C+C^3)} \right]^{\frac{1}{2}}, 
\end{gather}
which are consistent with all the symmetries. These are also consistent with the $F_4$ superpotentials \eqref{WF4f1} and \eqref{WF4f2} via the deformation of $\braket{B}=v^3$. 
These superpotentials are runaway and there is no stable supersymmetric vacuum for $1 \le F \le 3$. 
By introducing the non-perturbative superpotential from the KK-monopole \eqref{WsupKKE6}, which is available for all $F$, and by integrating out the Coulomb branch operator, we can reproduce the 4d effective superpotential for $F \le 4$ \cite{Cho:1997am}.

\begin{table}[H]\caption{3d $\mathcal{N}=2$ $E_6$ gauge theory with $4 \, \protect\Young[0]{1}$} 
\begin{center}
\scalebox{1}{
  \begin{tabular}{|c||c|c|c|c| } \hline
  &$E_6$&$SU(4)$&$U(1)$&$U(1)_R$  \\ \hline
$Q$&$\mathbf{27}$&${\tiny \yng(1)}$&1&$r$ \\ 
$\eta=\Lambda^b$&1&1&$6F$&$6F(r-1) +24=24r$  \\  \hline
$B:=Q^3$&1&${\tiny \yng(3)}$&3&$3r$ \\
$C:=Q^6$&1&${\tiny \overline{\yng(2)}}$&6&$6r$ \\
$D:=Q^{12}$&1&$1$&12&$12r$ \\ \hline
$Z_{SU(6)}$&1&1&$-6F=-24$&$-6F(r-1) -22 =2-24r$ \\ \hline
%$N_{5}:=Q^5$&1&${\tiny \overline{\yng(1)}}$&5&$5R_Q$ \\
%$N_{6}:=Q^6$&1&1&6& $6R_Q$ \\
%$N_{9}:=Q^5$&1&1&9& $9R_Q$  \\ \hline  
%$Z$&1&1&$-18$& $-18(R_Q-1) -16=2-18R_Q$  \\ \hline
  \end{tabular}}
  \end{center}\label{E640}
\end{table}

%%%%%%%%%%%%%%%%%%%%%%%%%%%%%%%%%%%%%%%%%%%%%%%%%%%%%%%%%%
\subsection{$E_6$ with two (anti-)fundamental flavors}
%%%%%%%%%%%%%%%%%%%%%%%%%%%%%%%%%%%%%%%%%%%%%%%%%%%%%%%%%%
In the $E_6$ gauge group, we can freely introduce fundamental and anti-fundamental quarks since there is no chiral and parity gauge anomalies in 4d and 3d, respectively. 
The next example is a 3d $\mathcal{N}=2$ $E_6$ gauge theory with two fundamental and two anti-fundamental matters, which is a ``vector-like'' theory. The corresponding 4d theory was studied in \cite{Grinstein:1998bu}. The Higgs branch operators are summarized in Table \ref{E622}. The Coulomb branch is described by a single operator $Z_{SU(6)}$. We will not explicitly write down the effective superpotential. In the superpotential, the Coulomb branch operator linearly couples to a homogeneous function of ($Q, \bar{Q})$ with degree $Q^{12} \bar{Q}^{12}$.

\begin{table}[H]\caption{3d $\mathcal{N}=2$ $E_6$ gauge theory with $2( \protect\Young[0]{1}+  \overline{\protect\Young[0]{1}})$} 
\begin{center}
\scalebox{1}{
  \begin{tabular}{|c||c|c|c|c|c|c| } \hline
  &$E_6$&$SU(2)$&$SU(2)$&$U(1)$&$U(1)$&$U(1)_R$  \\ \hline
$Q$&$\mathbf{27}$&${\tiny \yng(1)}$&1&1&0&$r$ \\
$\bar{Q}$&$\overline{\mathbf{27}}$&1&${\tiny \yng(1)}$&0&1&$\bar{r}$  \\
$\eta=\Lambda^b$&1&1&1&$6F$&$6\bar{F}$&$6F(r-1) +24=24r$  \\  \hline
$M_{Q\bar{Q}}:=Q\bar{Q}$&1&${\tiny \yng(1)}$&${\tiny \yng(1)}$&1&1&$r+\bar{r}$ \\
$B:=Q^3$&1&${\tiny \yng(3)}$&1&3&0&$3r$ \\
$\bar{B}:=\bar{Q}^3$&1&1&${\tiny \yng(3)}$&0&3&$3\bar{r}$ \\
$M_{2,2}:=Q^2 \bar{Q}^2$&1&${\tiny \yng(2)}$&${\tiny \yng(2)}$&2&2&$2r+2\bar{r}$ \\
$M_{1,4}:= Q\bar{Q}^4$&1&${\tiny \yng(1)}$&1&1&4&$r+4\bar{r}$\\
$M_{4,1}:=Q^4 \bar{Q}$&1&1&${\tiny \yng(1)}$&4&1&$4r +\bar{r}$ \\
$M_{3,3}:=Q^3 \bar{Q}^3$&1&${\tiny \yng(1)}$&${\tiny \yng(1)}$&3&3&$3r+3\bar{r}$ \\
$M_{4,4}:=Q^4 \bar{Q}^4$&1&1&1&4&4&$4r+4\bar{r}$ \\
$M_{6,6}:=Q^6 \bar{Q}^6$&1&1&1&6&6&$6r+6\bar{r}$\\ \hline
$Z_{SU(6)}$&1&1&1&$-12$&$-12$&$2-12r -12 \bar{r}$ \\ \hline
%$N_{5}:=Q^5$&1&${\tiny \overline{\yng(1)}}$&5&$5R_Q$ \\
%$N_{6}:=Q^6$&1&1&6& $6R_Q$ \\
%$N_{9}:=Q^5$&1&1&9& $9R_Q$  \\ \hline  
%$Z$&1&1&$-18$& $-18(R_Q-1) -16=2-18R_Q$  \\ \hline
  \end{tabular}}
  \end{center}\label{E622}
\end{table}

We can connect this low-energy dynamcis to the known s-confining phases. First, let us introduce a non-zero vev for $ \braket{M_{Q \bar{Q}}^{~~~22}}=v^2$, which breaks the gauge group into $Spin(10)$. The matter fields are decomposed into
\begin{align}
\mathbf{27} & \rightarrow \mathbf{16}+\mathbf{10}+\mathbf{1} \\
\overline{\mathbf{27}}& \rightarrow \overline{\mathbf{16}}+\mathbf{10}+\mathbf{1}
\end{align}
and a single set of a spinor and a conjugate spinor is eaten via the Higgs mechanism. As a result, we can flow to the 3d $\mathcal{N}=2$ $Spin(10)$ gauge theory with four vectors, one spinor and one conjugate spinor. This theory was studied in \cite{Nii:2018wwj} and it was shown that the low-energy dynamics is in s-confinement. The low-energy Coulomb branch operator is identified with $Z_{SU(6)} v^8$. 

Alternatively, by giving a non-zero vev to $\braket{B}:= \braket{f_{abc}Q^aQ^b Q^c}=v^3$, the low-energy theory becomes a 3d $\mathcal{N}=2$ $F_4$ gauge theory with three fundamental matters, which is again s-confining. The high- and low-energy Coulomb branches are related by $Z_{SU(6)} v^6 \sim Z_{USp(6)}$. This is consistent with our analysis of the Coulomb branches since $v^6$ is absorbing six fermion zero-modes of $Z_{SU(6)}$ and $Z_{SU(6)} v^6 $ has $18$ zero-modes.

%%%%%%%%%%%%%%%%%%%%%%%%%%%%%%%%%%%%%%%%%%%%%%%%%%%%%%%%%%
\subsection{$E_6$ with three fundamental and one anti-fundamental matters}
%%%%%%%%%%%%%%%%%%%%%%%%%%%%%%%%%%%%%%%%%%%%%%%%%%%%%%%%%%
The third example is a 3d $\mathcal{N}=2$ $E_6$ gauge theory with three fundamental matters and a single anti-fundamental matter, which is a ``chiral'' theory. The monopole operator corresponding to $Z_{SU(6)}$ has $18$ fundamental zero-modes and $6$ anti-fundamental zero-modes. The matter fields and the moduli coordinates are summarized in Table \ref{E631}. In the effective superpotential, the Coulomb branch operator $Z_{SU(6)}$ linearly couples to a homogeneous function of ($Q, \bar{Q})$ with degree $Q^{18} \bar{Q}^6$. At a generic point of the Higgs branch, the $E_6$ gauge symmetry could be completely higgsed. Hence, the dimension of the classical Higgs branch is $30$. The total number of the Higgs branch coordinates is $31$ and one constraint must be imposed between them. The F-flatness condition for $Z_{SU(6)}$ correctly reduces them by one. 

As in the previous subsection, by introducing a vev to $ \braket{M^{~~~31}_{Q \bar{Q}}}=v^2$, we can flow to the 3d $Spin(10)$ theory with four vectors and two spinors. As studied in \cite{Nii:2018wwj}, the low-energy dynamics of the $Spin(10)$ theory exhibits s-confinement. The low-energy Coulomb branch of the $Spin(10)$ theory is identified with $Z_{SU(6)} v^8$. Alternatively, when $B$ obtains a non-zero vev, the theory flows to the 3d $\mathcal{N}=2$ $F_4$ gauge theory with three fundamental matters, which was studied in the previous section and exhibits s-confinement. 

\begin{table}[H]\caption{3d $\mathcal{N}=2$ $E_6$ gauge theory with $3 \, \protect\Young[0]{1}+  \overline{\protect\Young[0]{1}}$} 
\begin{center}
\scalebox{0.95}{
  \begin{tabular}{|c||c|c|c|c|c| } \hline
  &$E_6$&$SU(3)$&$U(1)$&$U(1)$&$U(1)_R$  \\ \hline
$Q$&$\mathbf{27}$&${\tiny \yng(1)}$&1&0&$r$ \\ 
$\bar{Q}$&$\overline{\mathbf{27}}$&1&0&1&$\bar{r}$ \\ 
$\eta=\Lambda^b$&1&1&$18$&$6$&$18(r-1)+6(\bar{r}-1) +24=18r + 6 \bar{r}$  \\  \hline
$M_{Q \bar{Q}}:=Q \bar{Q}$&1&${\tiny \yng(1)}$&1&1&$r+ \bar{r}$ \\
$B:=Q^3$&1&${\tiny \yng(3)}$&3&0&$3r$ \\
$\bar{B}:= \bar{Q}^3$&1&1&0&3&$3 \bar{r}$ \\
$M_{2,2}:=Q^2 \bar{Q}^2$&1&${\tiny \yng(2)}$&2&2&$2r +2 \bar{r}$ \\
$M_{4,1}:=Q^4 \bar{Q}$&1&${\tiny \yng(2,2)}={\tiny \overline{\yng(2)}}$&4&1&$4r +\bar{r}$ \\
$M_{6,0}:=Q^6$&1&1&6&0&$6r$ \\
$M_{5,2}:=Q^5 \bar{Q}^2$&1&${\tiny \yng(2,2,1)}={\tiny \overline{\yng(1)}}$&5&2&$5r +2\bar{r}$ \\
$M_{9,3}:= Q^9 \bar{Q}^3$&1&1&9&3&$9r +3 \bar{r}$ \\ \hline
$Z_{SU(6)}$&1&1&$-18$&$-6$&$-18(r-1)-6(\bar{r}-1) -22=2-18r -6 \bar{r}$ \\ \hline 
%$N_{5}:=Q^5$&1&${\tiny \overline{\yng(1)}}$&5&$5R_Q$ \\
%$N_{6}:=Q^6$&1&1&6& $6R_Q$ \\
%$N_{9}:=Q^5$&1&1&9& $9R_Q$  \\ \hline  
%$Z$&1&1&$-18$& $-18(R_Q-1) -16=2-18R_Q$  \\ \hline
  \end{tabular}}
  \end{center}\label{E631}
\end{table}

%%%%%%%%%%%%%%%%%%%%%%%%%%%%%%%%%%%%%%%%%%%%%%%%%%%%%%%%%%
%%%%%%%%%%%%%%%%%%%%%%%%%%%%%%%%%%%%%%%%%%%%%%%%%%%%%%%%%%
\section{3d $\mathcal{N}=2$ $E_7$ gauge theory}
%%%%%%%%%%%%%%%%%%%%%%%%%%%%%%%%%%%%%%%%%%%%%%%%%%%%%%%%%%
%%%%%%%%%%%%%%%%%%%%%%%%%%%%%%%%%%%%%%%%%%%%%%%%%%%%%%%%%%
Finally, we investigate the 3d $\mathcal{N}=2$ $E_7$ gauge theory with $F$ fundamental matters. The fundamental representation is $\mathbf{56}$ dimensional whereas the adjoint representation is denoted by $\mathbf{133}$. The 3d s-confinement phase appears for $F=3$, where the index constraint \eqref{3dscon_ind} is satisfied. The Higgs branch\footnote{The Higgs branch for higher $F$ is not completely understood since there are very complicated syzygies between the Higgs branch coordinates. This makes the understanding of the exceptional Seiberg duality difficult even in 4d.} is studied in 4d \cite{Cho:1997am, Grinstein:1998bu, Pouliot:2001iw} and its coordinates for $F=3$ are summarized in Table \ref{E7}. We here focus on the analysis of the Coulomb branch in the 3d $\mathcal{N}=2$ $E_7$ gauge theory, which is absent in 4d. The Coulomb branch denoted by $Z_{SO(12)}$ corresponds to the gauge symmetry breaking 
\begin{align}
E_7 & \rightarrow so(12) \times u(1) \label{E7breaking} \\
\mathbf{56} & \rightarrow \mathbf{12}_{\pm 1} + \mathbf{32'}_{0} \\
\mathbf{133} & \rightarrow  \mathbf{66}_{0} + \mathbf{1}_{0} + \mathbf{1}_{\pm 2} + \mathbf{32}_{\pm 1}
\end{align}
where $Z_{SO(12)}$ corresponds to the unbroken $U(1)$ subgroup. 
This Coulomb branch was studied in \cite{Davies:2000nw} to study the 4d $\mathcal{N}=1$ $E_7$ pure SYM theory without a matter on a circle.
In this breaking, the massless component $\mathbf{32'}_{0}$ from the fundamental representations can make the low-energy vacuum of the $Spin(12)$ gauge theory stable and supersymmetric. Therefore, this branch can be exactly massless after including the quantum corrections. The mixed Chern-Simons terms are generated due to the one-loop diagrams of the massive components which have non-zero $U(1)$ charges. This leads to a non-trivial mixing between the $U(1)$ global and topological symmetries. The components $\mathbf{12}_{\pm 1}$ generate $12$ fermion zero-modes around the monopole background associated with $Z_{SO(12)}$. The massive components $\mathbf{1}_{\pm 2} + \mathbf{32}_{\pm 1}$ from the gaugino field have the $34$ adjoint zero-modes around the monopole. These zero-modes give rise to the non-zero charges to the Coulomb branch operator $Z_{SO(12)}$ as in Table \ref{E7}. Although we will not explicitly write down the effective superpotential of the s-confining phase for $F=3$, the Coulomb branch $Z_{SO(12)}$ linearly couples to a homogeneous function of the Higgs branch operators with degree $Q^{36}$ in the superpotential.

By flowing to the Higgs branch, we can find the consistency of the Coulomb branch introduced above. When a single fundamental matter obtains a non-zero vacuum expectation value such that $\braket{B_4}$ is non-zero, the $E_7$ gauge group is higgsed into the $E_6$ gauge group. The fundamental $\mathbf{56}$ matter reduces to the (anti-)fundamental flavors $\mathbf{27}+\overline{\mathbf{27}}$ (plus two singlets) and the Higgs mechanism eats a single set of $\mathbf{27}+\overline{\mathbf{27}}$. Therefore, the theory flows to the 3d $\mathcal{N}=2$ $E_6$ gauge theory with two (anti-)fundamental flavors. This theory was studied in the previous section, which exhibited the s-confinement phase. From a viewpoint of the above breaking \eqref{E7breaking}, giving a vev to $\mathbf{56}$ corresponds to a vev for $ \mathbf{32'}_{0}$, which breaks $SO(12)$ into $SU(6)$. This is consistent with the Coulomb branch $Z_{SU(6)}$ of the $E_6$ theory. The two sets of massive vectors $\mathbf{12}_{\pm 1} $ (one set of $\mathbf{12}_{\pm 1}$ is eaten via the Higgs mechanism.) create $24$ fermion zero-modes and exactly the same as the zero-modes of the $E_6$ Coulomb branch. The low- and high-energy Coulomb branches are related as 
\begin{align}
Z_{SO(12)} v^{12} \sim Z_{SU(6)},
\end{align}
where $v$ is a vev for the fundamental matter $\mathbf{56}$. We can introduce another expectation value to $M_2$ (a composite constructed from an anti-symmetric invariant tensor $f^{\alpha \beta}$), where the theory flows to the 3d $\mathcal{N}=2$ $Spin(11)$ theory with five vectors and one spinor. As studied in \cite{Nii:2018wwj}, the low-energy dynamics of the $Spin(11)$ theory exhibits s-confinement. This is another test of our analysis.

\begin{table}[H]\caption{3d $\mathcal{N}=2$ $E_7$ gauge theory for $F=3$} 
\begin{center}
\scalebox{1}{
  \begin{tabular}{|c||c|c|c|c| } \hline
  &$E_7$&$SU(3)$&$U(1)$&$U(1)_R$  \\ \hline
$Q$&$\mathbf{56}$&${\tiny \yng(1)}$&1&$r$ \\
$\lambda$&$\mathbf{133}$&1&0&$1$ \\
$\eta=\Lambda^b$&1&1&36&$36r$  \\  \hline
$M_2:=Q^2$&1&${\tiny \overline{\yng(1)}}$&2&$2r$ \\
$B_4:=Q^4$&1&${\tiny \yng(4)}$&4&$4r$ \\
$B_6:=Q^6$&1&${\tiny \overline{\yng(3)}}$&6&$6r$ \\
$B_8:=Q^{8}$&1&${\tiny \yng(2)}$&8&$8r$ \\ 
$B_{12}:=Q^{12}$&1&1&12&$12r$ \\
$B_{18}:=Q^{18}$&1&1&18&$18r$ \\ \hline
%$N_{5}:=Q^5$&1&${\tiny \overline{\yng(1)}}$&5&$5R_Q$ \\
%$N_{6}:=Q^6$&1&1&6& $6R_Q$ \\
%$N_{9}:=Q^5$&1&1&9& $9R_Q$  \\ \hline  
$Z_{SO(12)}$&1&1&$-12F=-36$& $-12F(r-1) -34=2-36r$  \\ \hline
  \end{tabular}}
  \end{center}\label{E7}
\end{table}

%[4d-3d relation]
 We can connect the 3d theory to the 4d $\mathcal{N}=1$ $E_7$ gauge theory with three fundamental matters. The 4d low-energy dynamics exhibits a quantum-deformed moduli space with a single constraint \cite{Cho:1997am, Grinstein:1998bu}. By putting the 4d theory on a circle-compactified manifold $\mathbb{S}^1 \times \mathbb{R}^3$, we can connect the low-energy 3d and 4d dynamics. When the 4d theory is defined on a circle, we have an additional monopole configuration known as a KK-monopole (or called a twisted instanton) which generates a non-perturbative superpotential \cite{Lee:1997vp, Aharony:1997bx, Davies:2000nw}
\begin{align}
W= \eta Z, 
\end{align}
where we defined $\eta:= \Lambda^b$. $\Lambda$ is a dynamical scale of the 4d gauge interaction and $b$ is a coefficient of the one-loop beta function. By adding this potential to the 3d s-confinement phase and integrating over the Coulomb branch operator which is absent in a 4d limit, we can recover the 4d deformed moduli space studied in \cite{Cho:1997am, Grinstein:1998bu}.

%%%%%%%%%%%%%%%%%%%%%%%%%%%%%%%%%%%%%%%%%%%%%%%%%%%%%%%%%%
%%%%%%%%%%%%%%%%%%%%%%%%%%%%%%%%%%%%%%%%%%%%%%%%%%%%%%%%%%
\section{Summary and Discussion}
%%%%%%%%%%%%%%%%%%%%%%%%%%%%%%%%%%%%%%%%%%%%%%%%%%%%%%%%%%
%%%%%%%%%%%%%%%%%%%%%%%%%%%%%%%%%%%%%%%%%%%%%%%%%%%%%%%%%%
%[What we have done]
In this paper, we found the s-confinement phases for the 3d $\mathcal{N}=2$ exceptional gauge theories with fundamental matters by identifying the quantum Coulomb branch operators which are stable and exactly massless. We observed that the s-confinement phases appear when the index constraint \eqref{3dscon_ind} is satisfied. Those new examples include $F_4$, $E_6$ and $E_7$ gauge groups, where the Coulomb branch operators preserve $USp(6)$, $SU(6)$ and $SO(12)$ gauge symmetries, respectively. We argued that the Coulomb branch operators discussed here are connected to each other via the Higgs branch flow. 
By giving a non-zero vev to the other Higgs branch coordinate, we can also flow to the known s-confinement phases of the 3d $\mathcal{N}=2$ $Spin(N)$ gauge theories \cite{Nii:2018wwj}. For the $E_6$ cases, we studied the ``chiral'' and ``vector-like'' theories with fundamental and anti-fundamental matters. By introducing the non-perturbative superpotential from the KK-monopole (a twisted instanton on $\mathbb{S}^1 \times \mathbb{R}^3$), we reproduced the quantum-deformed moduli space of the corresponding 4d theories.

%[Future work]

%[Exceptional gauge theory with many flavors]
In this paper, we focused on the s-confinement phases of the exceptional gauge theories. Therefore, the number of the fundamental flavors was restricted from above. For the theory with more flavors, it will be anticipated that the theory has a non-abelian Coulomb phase and hence the moduli space will include singularities which signal that there are additional massless degrees of freedom along the singularities. As we briefly explained in the $F_4$ case, the Coulomb branch for the theory with more fundamental matters can have additional Coulomb branch and the low-energy analysis will become more complicated. It would be important to investigate those phases. This additional Coulomb branch is consistent with the analysis of the 3d $Spin(N)$ gauge theories \cite{Nii:2018wwj} where we need to introduce various Coulomb branch operators depending on the number of spinor matters. As a further check of this, we can compute the superconformal indices \cite{Bhattacharya:2008bja, Kim:2009wb, Kapustin:2011jm} by employing a localization technique. This will be left as a future work. 

%[exceptional dualities]
For the exceptional gauge groups, the self-dualities were proposed in 4d \cite{Pouliot:1995zc, Ramond:1996ku, Distler:1996ub, Karch:1997jp} although they do not satisfy the 't Hooft anomaly matching conditions of the discrete symmetries except for the $G_2$ case \cite{Csaki:1997aw}. The resolution of this mismatch is to consider the Kutasov-type duality \cite{Kutasov:1995ve, Kutasov:1995np, Csaki:1997aw} which is a Seiberg duality with a tree-level superpotential breaking the discrete symmetries. However, it is not clear whether or not this way out completely cures the problem. It is worthwhile studying the 3d and 4d dualities for the exceptional gauge theories and more rigorously test the validity of the proposed dualities. 

%[SCIを見るべき]

%%%%%%%%%%%%%%%%%%%%%%%%%%%%%%%%%%%%%%%%%%%%%%%%%%%%%%%%%%
%%%%%%%%%%%%%%%%%%%%%%%%%%%%%%%%%%%%%%%%%%%%%%%%%%%%%%%%%%
\section*{Acknowledgments}
%%%%%%%%%%%%%%%%%%%%%%%%%%%%%%%%%%%%%%%%%%%%%%%%%%%%%%%%%%
%%%%%%%%%%%%%%%%%%%%%%%%%%%%%%%%%%%%%%%%%%%%%%%%%%%%%%%%%%
This work is supported by the Swiss National Science Foundation (SNF) under grant number PP00P2\_183718/1.

%%%%%%%%%%%%%%%%%%%%%%%%%%%%%%%%%%%%%%%%%%%%%%%%%%%%%%%%%%
%%%%%%%%%%%%%%%%%%%%%%%%%%%%%%%%%%%%%%%%%%%%%%%%%%%%%%%%%%

%\bibliographystyle{unsrt}
\bibliographystyle{ieeetr}
\bibliography{exceptional}

\end{document}